# Estimating sample size in dental research

Hoi-Jeong Lim

*Department of Orthodontics, Dental Science Research Institute*
*Chonnam National University School of Dentistry*
*Gwangju, South Korea*

*Abstract*— **Determination of sample size is critical, however not easy to do. Sample size defined as the number of observations in a sample should be big enough to have a high likelihood of detecting a true difference between groups. Practical procedure for determining sample size, using G\*power and previous dental articles, is shown in this study. Examples involving independent t-test, paired t-test, one-way analysis of variance(ANOVA), and one-way repeated-measures(RM) ANOVA are used. The purpose of this study is to enable researchers with non-statistical backgrounds to use in practice freely available statistical software G\*power to determine sample size and power.**

## I. Introduction

Sample size refers to the number of patients, individuals, or animals participating in a study. The purpose of calculating sample size in clinical research is to determine the appropriate sample size needed to yield significant research results. If a study is conducted with a sample that is too small, it may have low statistical power, leading to insufficient evidence to draw conclusions or resulting in inaccurate conclusions. Conversely, conducting a study with too large a sample can waste time and resources, lead to loss of follow-up during the observation period, and raise ethical issues. Additionally, while statistical significance may be achieved, the results may not be clinically meaningful. Designing a study with an inappropriate sample size can lead to incorrect conclusions and potentially inappropriate treatments. Therefore, calculating the sample size during the research planning stage is an essential step.

## II. Factors to consider for sample size calculation

There are many factors to consider when determining power and sample size, but important considerations include Type I error, Type II error, and effect size. Each statistical method has its own formula for calculating sample size, so it is essential to carefully plan which variable will be the primary endpoint and which statistical method will be used for analysis during the research planning stage. Let's take a closer look at the factors to consider below.

### A. Type I error (α), Type II error (β), and Statistical power

Hypothesis testing is based on samples, so there is always a possibility of error. These errors include Type I error and Type II error. Type I error is the probability of claiming that there is an effect when there is none, while Type II error is the probability of claiming that there is no effect when there is one. Therefore, the power of a test is the probability of correctly identifying an effect when it exists. Typically, a fixed value of 0.05 is used for Type I error and 0.8 for power. Thus, while the concept of error is difficult, it is easy to use.

$$\text{Type I errors} = P[Reject\ H_0 | H_0 : True]$$

$$\text{Type II errors} = P[Do\ not\ reject\ H_0 | H_0 : False]$$

$$\text{Statistical Power} = 1 - \beta = P[Reject\ H_0 | H_0 : False]$$

$$\text{where } H_0 : \text{Null hypothesis}$$

### B. Effect size

Effect size refers to the minimum difference that is clinically meaningful. Determining this effect size is also a goal of clinical trials. As the effect size increases, the power of the test increases, but the sample size decreases. Conversely, if the effect size is small, the sample size must increase in order to significantly detect that small difference. Effect size is obtained from means, standard deviations, proportions, etc., provided in previous studies or pilot studies.

### C. Types of Hypothesis Testing

A two-tailed test is used to determine whether the means of two groups are the same or different under the null hypothesis and the alternative hypothesis, while a one-tailed test checks whether the mean of one group is greater than or less than the mean of the other group.

### D. Drop rate

Typically, the drop rate is set to less than 20%. In most cases, a drop rate of 10% is used, but if the study period is extended, more follow-up losses may occur, allowing for a slightly higher drop rate. In this case, the N value calculated using G\*power is divided by 0.9 to obtain the Final N (Final N: N = 1 : 0.9).

### E. Determination of the primary endpoint

In the research planning stage, the primary endpoint must be determined, and the sample size should be calculated based on that variable. The primary endpoint is the main outcome measured at the end of the study to see if the given treatment is effective. For example, it refers to the number of deaths from the disease being studied or the difference in the number of survivors between the treatment group and the control group.

## III. Calculating sample size using G\*power

G\*power was created by the University of Düsseldorf in Germany, and you can download G\*power 3.1.9.2 for free at http://www.gpower.hhu.de/en.tml. Below, we discuss how to calculate statistical power and sample size using G\*power based on papers published in dental journals according to different statistical methods.

### A. Sample size calculation using the independent t-test

Based on "Table 1" used in the paper by Miyawaki et al. published in the 2004 American Journal of Orthodontics[1], the power and sample size were calculated. This "Table 1" tested the hypothesis of whether the use of an occlusal splint increases salivary flow rate during relaxation, clenching, and chewing in both bruxism patients and a normal group. As a result, a comparison of salivary flow rates between 8 bruxism patients and 8 normal individuals did not yield significant differences across all variables. In this case, to determine whether the power exceeds 80% with a sample size of 8, the standard deviation (SD) was calculated using the standard error (SE) provided in the table, as shown in the equation below.

| splint/test movements | Control group (n = 8) | | | Bruxism group (n = 8) | | | | Power+ | Min N# | Final N§ |
| --- | --- | --- | --- | --- | --- | --- | --- | --- | --- | --- |
| | Mean | SE | SD* | Mean | SE | SD* | P | | ( per gp ) | ( per gp ) |
| Without splint | | | | | | | | | | |
| Relaxing | 0.49 | 0.05 | 0.1414 | 0.42 | 0.07 | 0.1980 | ns | 0.1182 | 96 | 107 |
| Chewing-like movement | 0.66 | 0.05 | 0.1414 | 0.49 | 0.11 | 0.3111 | ns | 0.2589 | 33 | 37 |
| Clenching | 0.52 | 0.05 | 0.1414 | 0.37 | 0.07 | 0.1980 | ns | 0.3688 | 22 | 25 |
| With splint | | | | | | | | | | |
| Relaxing | 0.64 | 0.05 | 0.1414 | 0.52 | 0.14 | 0.3960 | ns | 0.1171 | 98 | 109 |
| Chewing-like movement | 0.89 | 0.08 | 0.2263 | 0.81 | 0.21 | 0.5940 | ns | 0.0627 | 497 | 553 |
| Clenching | 0.63 | 0.05 | 0.1414 | 0.64 | 0.15 | 0.4243 | ns | 0.0504 | 15701 | 17446 |

*SE*, Standard error; *ns*, not significant between groups (unpaired *t* test or Mann-Whitney *U* test between groups).
*SD; standard deviation
+Power; probability that it rejects a false null hypothesis.
≠Min N; minimum sample size calculated by G*Power
§Final N; Min N considering the drop rate 10%

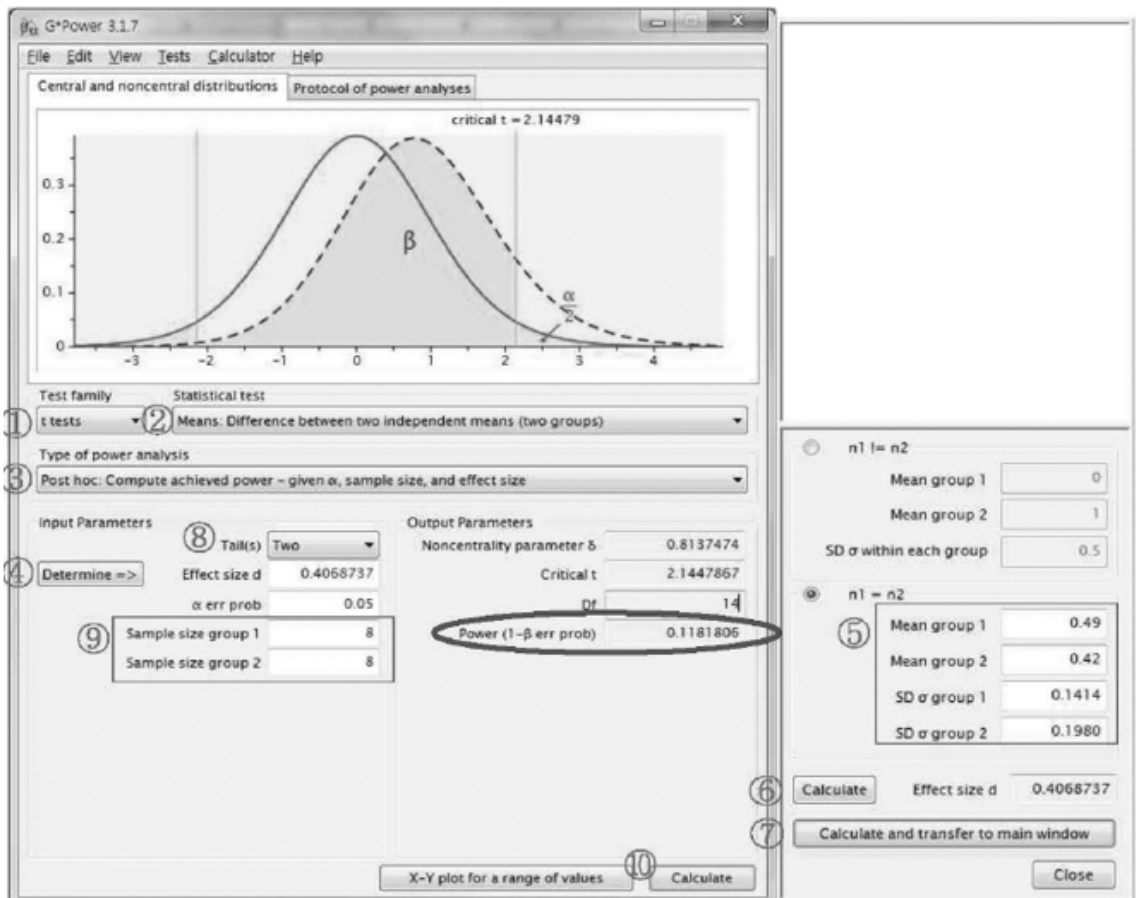

① Select. t tests from the test family
② To choose the independent t-test, select Means: Difference between two independent means.
③ To calculate the statistical power, select Post hoc. To calculate the sample size, select A priori.
④ Click on Determine to calculate the effect size.
⑤ Move to the next window and input the means and standard deviation of the two groups under the condition that sample sizes are the same. (n1=n2)
⑥ Click the Calculate button to compute the effect size.
⑦ Click Calculate and transfer to main window to transfer the calculated effect size to Effect size d in the window next to it.
⑧ Select Two from Tail(s) to choose a two-tailed test. (One: refers to a one-tailed test.)
⑨ Input the sample size of 8 for both groups.
⑩ Click the Calculate button to calculate the Statistical Power.

Fig. 1.   Power analysis based on independent t-test

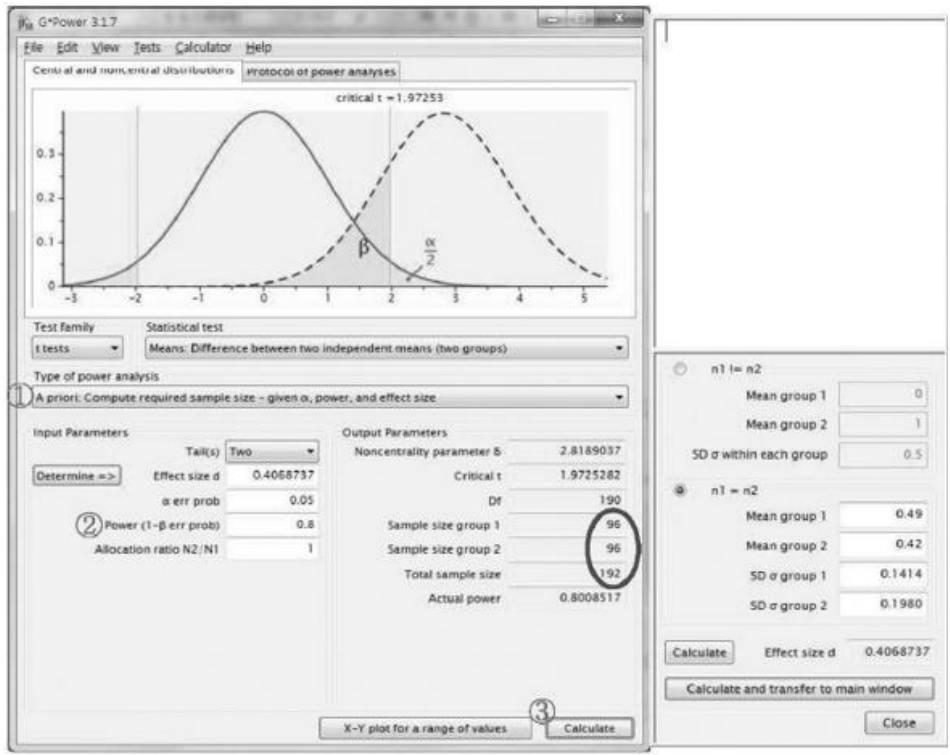

The rest is the same as the post hoc case mentioned above.
① In Figure 1, change post hoc in the type of power analysis to A priori.
② Change Power to 0.8.
③ Press the Calculate button to obtain the sample size of 96 for each group.

Fig. 2. Sample size determination based on independent t-test

$$SE = \frac{SD}{\sqrt{n}}$$

Using G*power, the results of the power analysis for items 1 to 10 below Figure 1 showed that whether using the sprint or not, the power was low, all below 40%.

The study concluded that there was no significant difference between the two groups based on the results of an independent t-test with a sample size of 8 per group. However, to achieve 80% power, it was found that the sample size needed for the variable "Relaxing without splint" is 96 per group, as determined by G*power. Considering a drop rate of 10%, the sample size was calculated by dividing 96 by 0.9, resulting in a total sample size of 107. If the main outcome variable in this study is "Relaxing without splint," at least 107 participants should be recruited per group "Table 1".

### B. Sample size calculation using paired t-test

In a paper written by Benson et al. published in the Angle Orthodontist journal in 2005[2], the difference in the demineralized lesion area between Captured Slides and Digital Camera Images was measured, and a significant difference was found between the two images (p=0.029). Additionally, there was a significant difference in the average gray level of the lesion area compared to the average gray level of the healthy area (p=0.002, p=0.001). However, no significant difference was found in luminance intensity (LI)% between the two images (p=0.148). The power of the test was calculated to determine if the sample size for this table was appropriate.

The results of a paired t-test conducted with a sample size of 27 sites concluded that there is a significant difference between the two images in terms of Area (p=0.029), but the power obtained was less than 0.8 (power=0.62). This means that just because the p-value is less than 0.05, it does not imply that the power is greater than 0.8, so it is necessary to calculate the post-hoc power in this case. In other words, a p-value less than 0.05 does not necessarily indicate that a sufficient sample size has been achieved. To ensure that the p-value is less than 0.05 while having a power greater than 0.8, at least 41 research subjects need to be included in the study. Referring to "Fig. 5", when p>0.05, the sample size is less than 22, and even if it exceeds 22, the power remains below 0.8 up to a sample size of 40. In the Lesion and Sound variables shown in the table below, the power exceeded 0.9, and the Lesion N=27 and Sound N=17 shown in the upper table indicate that they secured the minimum sample size (N=24, 14) needed for 80% power. However, for the LI% variable, a sample size of 109 research subjects must be secured, considering a 10% drop rate, to find a significant difference "Table 2, Table 3, Fig. 5".

### C. One-way ANOVA

The study by Aslan published in the Angle Orthodontist journal in 2014[4] aimed to evaluate the dentofacial effects by comparing the average measurements of Skeletal, Dental, and Soft Tissue Parameters among three groups: (1) the group using miniscrew anchorage with the Forsus$_{TM}$ Fatigue Resistant Device (FRDMS), (2) the conventional FRD group, and (3) the untreated class II control group. As a result, the measurements of U1/HRL and L1-VRL showed significant differences among the three groups.

TABLE II. POWER AND SAMPLE SIZE DETERMINATION BASED ON THE TABLE4 OF THE STUDY2) THAT USED PAIRED T-TEST

| | Paired Differences | | 95% CI of the Difference | | | | | | |
| | Mean | SD | Lower | Upper | P | N | Power+ | Min N≠ | Final N§ |
|---|---|---|---|---|---|---|---|---|---|
| Area | -0.29 | 0.64 | -0.54 | -0.03 | 0.029 | 27 | **0.6206** | **41** | **46** |
| Lesion | 9.08 | 13.77 | 3.63 | 14.52 | 0.002 | 27 | **0.9094** | **21** | **24** |
| Sound | 12.11 | 12.93 | 5.46 | 18.76 | 0.001 | 17 | **0.9517** | **12** | **14** |
| LI% | -2.86 | 9.96 | -6.80 | 1.08 | 0.148 | 27 | **0.3009** | **98** | **109** |

+Power; probability that it rejects a false null hypothesis.
≠Min N; minimum sample size calculated by G*Power
§Final N; Min N considering the drop rate 10%

TABLE III. THE RELATIONSHIP AMONG P-VALUE, POWER, AND SAMPLE SIZE IN DEMINERALIZED LESION AREA

| p>0.05 and power<0.8 | p<0.05 and power<0.8 | p<0.05 and power>0.8 |
|---|---|---|
| <22 | 22≤ <41 | ≥41 |

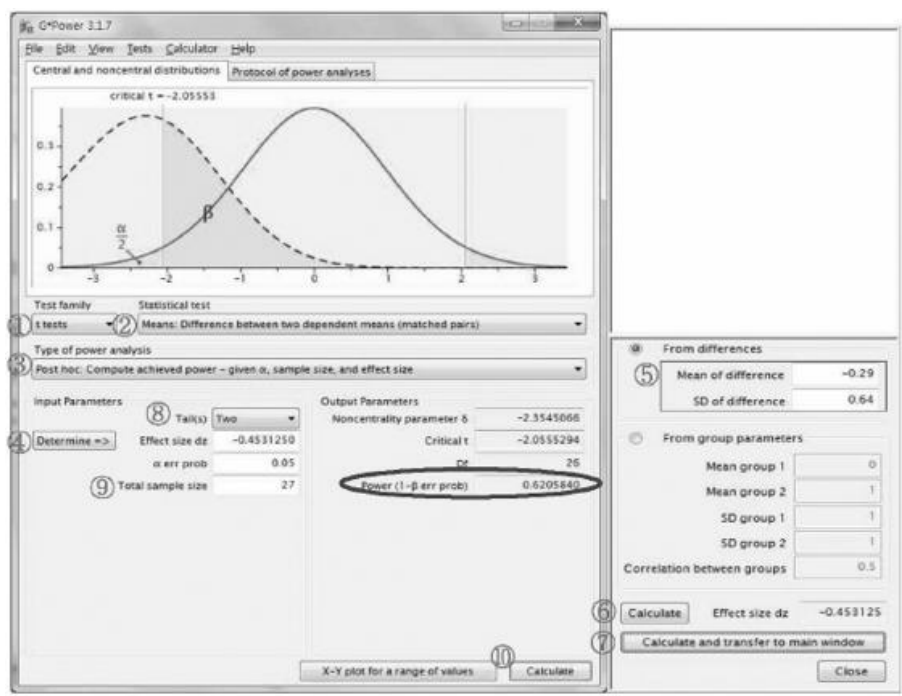

① Select t tests from the test family.
② To select a paired t-test, select Means: Difference between two dependent means.
③ To calculate the statistical power, select Post hoc. To calculate the sample size, select A priori.
④ Click Determine to calculate the effect size.
⑤ Move to the window next to it and enter the mean and standard deviation of the differences in From differences.
⑥ Clicking the Calculate button will calculate the effect size.
⑦ Click Calculate and transfer to main window to transfer the calculated effect size to Effect size d in the window next to it.
⑧ By selecting Two from Tail(s), choose a two-tailed test. (One: refers to a one-tailed test.)
⑨ Enter the total sample size as 27.
⑩ Click the Calculate button to calculate the statistical Power.

Fig. 3. Power analysis based on paired t-test

The results of the power analysis for the U1/HRL and L1-VRL variables in the table above showed that the power was greater than 0.8, indicating that the minimum required sample sizes of 15 and 45 have been secured. However, for other variables, the power was less than 0.8, and a larger sample size was needed to find significance between groups. If the primary outcome variable is U1/HRL or L1-VRL, there is no issue with the selected sample size, but for other variables, a larger sample size should be chosen to begin the study "Table 4".

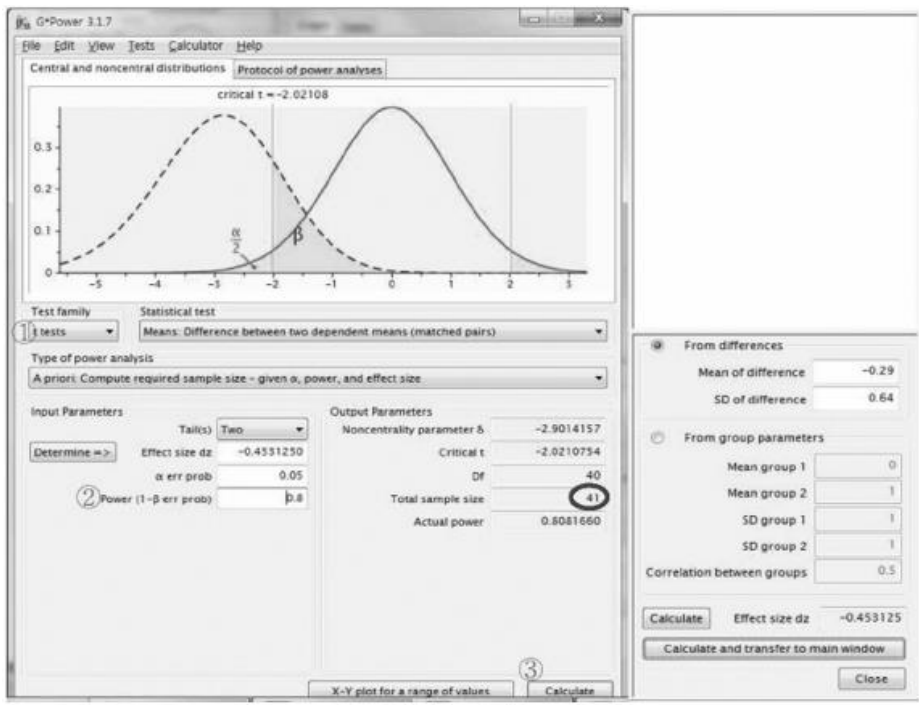

① In Figure 3, change Post hoc to A priori in the type of power analysis.
② Change Power to 0.8.
③ Click the Calculate button to obtain the sample size of 41 for each group.

Fig. 4.  Sample size determination based on paired t-test

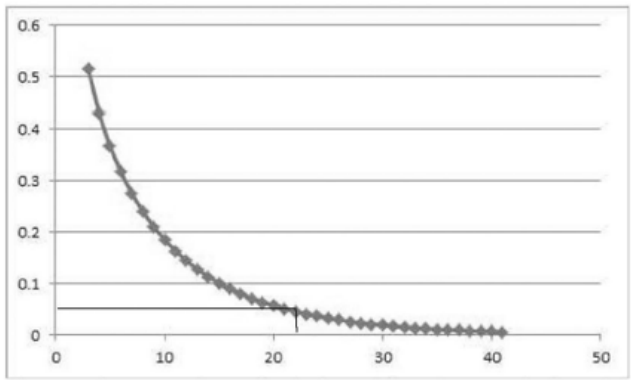

① Calculate the test statistic $T = \dfrac{\overline{D}}{S_D / \sqrt{N}}$ and degrees of freedom (N-1) for the paired t-test with sample sizes(N) ranging from 3 to 41.

② Calculate the statistical power using the sample size and the effect size obtained from Figure 3.

③ Create a graph with the p-value on the y-axis and the sample size on the x-axis.

④ We found that the sample size for which the p-value is greater than 0.05 is when it is larger than 22.

Fig. 5.  Plot for total sample size(x-axis) against p-value(y-axis)

TABLE IV.    POWER AND SAMPLE SIZE DETERMINATION BASED ON THE TABLE2 OF THE STUDY[4] THAT USED ONE-WAY ANOVA

|  | 1 ( n = 16) | 2 ( n = 17) | 3 ( n = 15) | P | SD⁺ | Power⁻ | Min N* | Final N§ |
|---|---|---|---|---|---|---|---|---|
| SN, mm | 74.70 ± 3.70 | 76.87 ± 3.13 | 74.66 ± 3.46 | .082 | 3.43 | 0.4323 | 108 | 120 |
| SNA, ° | 80.53 ± 3.76 | 78.27 ± 2.19 | 78.14 ± 3.02 | .067 | 3.04 | 0.5709 | 78 | 87 |
| A-VRL, mm | 70.03 ± 3.50 | 72.32 ± 3.72 | 69.54 ± 3.29 | .079 | 3.52 | 0.5395 | 84 | 94 |
| U1/HRL, ° | 112.03 ± 5.11 | 110.17 ± 5.31 | 95.55 ± 9.62 | <.001* | 6.89 | 0.9999 | 15 | 17 |
| L1-VRL, mm | 68.00 ± 3.98 | 68.91 ± 4.66 | 64.10 ± 4.14 | <.001* | 4.28 | 0.8309 | 45 | 50 |
| L1-MP, mm | 43.26 ± 2.24 | 43.01 ± 2.47 | 42.14 ± 2.87 | .057 | 2.53 | 0.1836 | 282 | 314 |

₊ FRD indicates fatigue resistant device; FRDMS, FRD treatment used with miniscrew anchorage.
Values for groups 1, 2, and 3 are expressed as the mean ± the standard deviation
* P < .01; one-way analysis of variance.
*SD; standard deviation
+power; probability that it rejects a false null hyphothesis
≠Min N; minimum sample size calculated by G*Power
§Final N; Min N considering the drop rate 10%

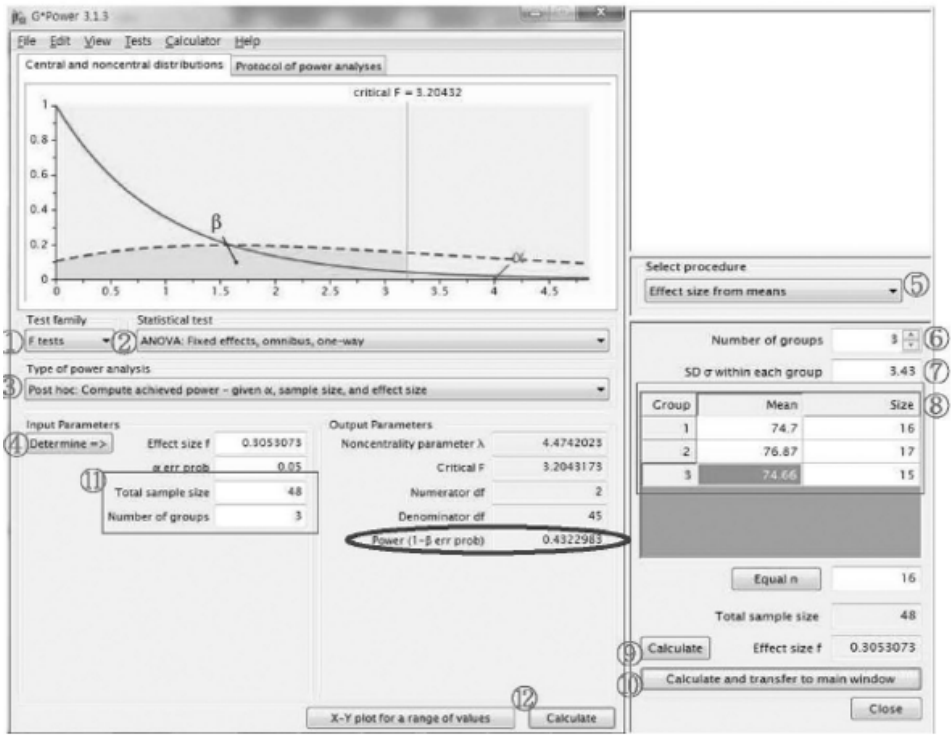

① Select F tests from test family.

② To select One ANOVA, select ANOVA: Fixed effects, omnibus, one-way.

③ To calculate the statistical power, select Post hoc. To calculate the sample size, select A priori.

④ Click Determine to calculate the effect size.

⑤ Move to the window next to it and select Effect size from means in Select procedure.

⑥ Enter the number of groups.

⑦ Calculate the standard deviation within the group as follows.

$$s_p = \sqrt{\frac{(n_1-1)s_1^2 + (n_2-1)s_2^2 + \cdots + (n_k-1)s_k^2}{N-k}} \quad (pooled\ estimate)$$

$n_1, n_2, \ldots, n_k$ : Sample size of each group

$s_1, s_2, \ldots, s_k$ : Standard deviation of each group

$N = n_1 + n_2 + \ldots + n_k$ : total sample size, $k$ = Number of group

⑧ Enter the mean and sample size for each group.

⑨ Click the Calculate button to compute the total sample size and effect size.

⑩ Click Calculate and transfer to main window to transfer the calculated effect size to Effect size d in the window next to it.

⑪ Enter the total sample size and number of groups.

⑫ Click the Calculate button to calculate the statistical Power.

Fig. 6. Power analysis based on one-way ANOVA

TABLE V. POWER AND SAMPLE SIZE DETERMINATION BASED ON THE TABLE2 OF THE STUDY[5] THAT USED ONE-WAY REPEATED-MEASURES ANOVA

| | Presurgery (T0) | Postsurgery (T1) | Follow-up (T2) | P-value | Power+ | Min N* | Final N§ |
|---|---|---|---|---|---|---|---|
| Axial condylar angle (right) | 74.90 ± 6.88a | 68.58 ± 7.46b | 68.68 ± 7.16b | .048* | 0.9986 | 12 | 14 |
| Axial condylar angle (left) | 73.24 ± 5.27a | 67.44 ± 6.40b | 67.13 ± 6.09b | .020* | 0.9995 | 11 | 13 |
| Anterior space (right) | 1.81 ± 0.73a | 2.70 ± 0.57 | 1.88 ± 0.42a | .043* | 0.9338 | 20 | 23 |
| Superior space (right) | 2.67 ± 0.79 | 2.28 ± 0.64 | 2.42 ± 0.51 | .464 | 0.2076 | 99 | 110 |
| Posterior space (right) | 2.43 ± 0.65 | 68.91 ± 4.66 | 2.25 ± 0.46 | .703 | 0.3181 | 66 | 74 |

Distances are given in mm, angles in degrees.

FH, Frankfort horizontal.

*Significant difference in the 3 groups by analysis of variance ($P < .05$).

a,b,c The same superscripts indicate no statistically significant differences among the indicated groups ($P > .05$).

+Power; probability that it rejects a false null hypothesis.

≠Min N; minimum sample size calculated by G*Power

§Final N; Min N considering the drop rate 10%

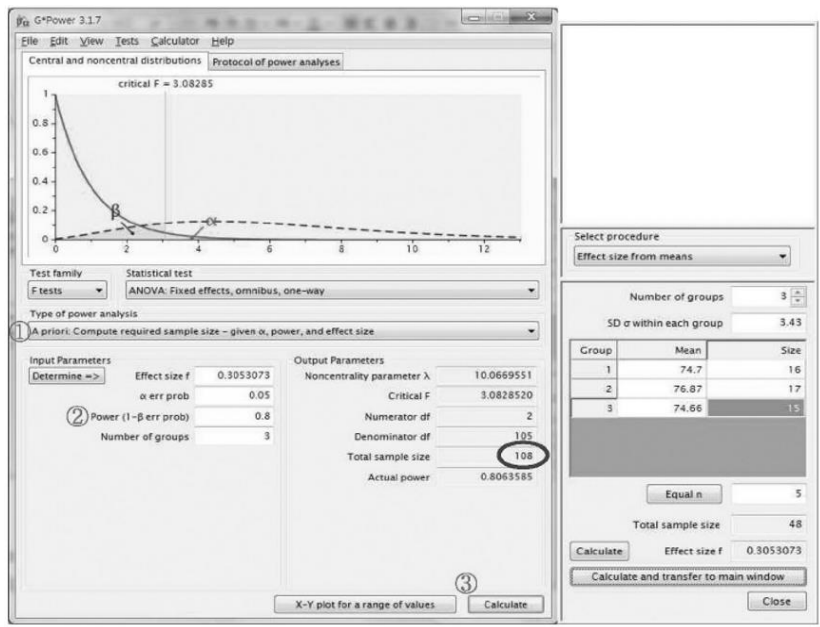

① In Figure 6, change Post hoc to A priori in Type of power analysis.

② Change Power to 0.8.

③ Click the Calculate button to calculate the total sample size. (108/3 = 36 people are required per group)

Fig. 7.   Sample size determination based on one-way ANOVA

### Mauchly's Test of Sphericity[a]

Measure : MEASURE_1

| Within Subjects Effect | Mauchly's W | Approx. Chi-Square | df | Sig. | Epsilon[b] | | |
|---|---|---|---|---|---|---|---|
| | | | | | Greenhouse-Geisser | Huynh-Feldt | Lower-bound |
| superior_space | .814 | 4.948 | 2 | .084 | .843 | .897 | .500 |

Tests the null hypothesis that the error covariance matrix of the orthonormalized transformed dependent variables is proportional to an identity matrix.

a. Design: Intercept + Group
  Within Subjects Design: session

b. May be used to adjust the degrees of freedom for the averaged tests of significance. Corrected tests are displayed in the Tests of Within-Subjects Effects table.

Fig. 8.   Mauchly's sphericity test in SPSS output

### Tests of Within-Subjects Effects

측도: MEASURE_1

Variance explained by effect

| Source | | Type III Sum of Squares | df | Mean Square | F | P-value |
|---|---|---|---|---|---|---|
| superior_space | Sphericity Assumed | 2.331 | 2 | 1.166 | 1.939 | .155 |
| | Greenhouse-Geisser | 2.331 | 1.686 | 1.383 | 1.939 | .162 |
| | Huynh-Feldt | 2.331 | 1.794 | 1.299 | 1.939 | .160 |
| | Lower-bound | 2.331 | 1.000 | 2.331 | 1.939 | .176 |
| 오차(superior_space) | Sphericity Assumed | 30.059 | 50 | .601 | | |
| | Greenhouse-Geisser | 30.059 | 42.148 | .713 | | |
| | Huynh-Feldt | 30.059 | 44.859 | .670 | | |
| | Lower-bound | 30.059 | 25.000 | 1.202 | | |

Error Variance

Fig. 9.   Test of Within-Subjects Effects in SPSS output

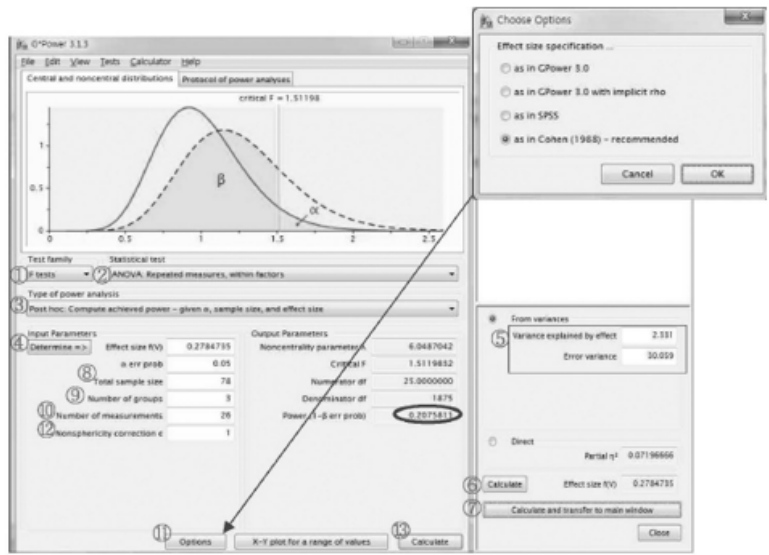

① Select F tests from the test family.

② To select One RM ANOVA, select ANOVA: Repeated measures, within factors.

③ To calculate the power, select Post hoc. To calculate the sample size, select A priori.

④ Click Determine to calculate the effect size.

⑤ Move to the window next to it and enter SSTrt in Variance explained by effect and SSE in Error variance in From variances by looking at the SPSS output.

⑥ Click Calculate to calculate the Effect size.

⑦ Click Calculate and transfer to main window to copy the Effect size to the main window.

⑧ Enter the total sample size.

⑨ Enter the number of groups.

⑩ Enter the number of measurements for each group.

⑪ Click the Options button, select as in Cohen(1988)-recommend, and click the OK button.

⑫ Since sphericity is satisfied, the nonsphericity correction ε value is entered as 1.

⑬ Click the Calculate button to calculate the Power.

Fig. 10. Power analysis based on one-way repeated-measures ANOVA

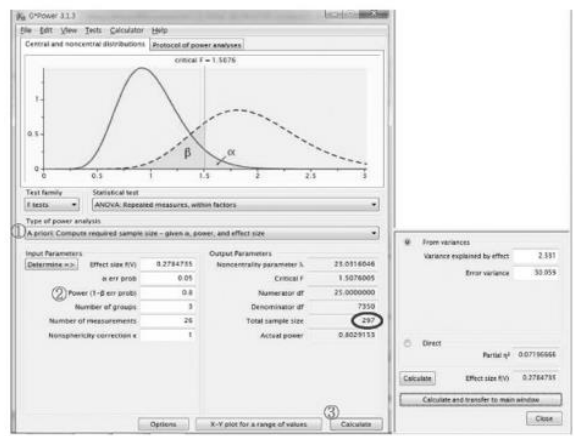

① In Figure 10, change Post hoc to A priori in Type of power analysis.
② Change Power to 0.8.
③ Click the Calculate button to calculate the sample size for each group. (297/3=99 per time point)

Fig. 11. Sample size determination based on one-way repeated-measures ANOVA

### D. One-way repeated-measures(RM) ANOVA

The study by Kim et al. published in the journal Oral Surg Oral Med Oral Pathol Oral Radiol Endod in 2011[5] compares the condylar axis and anteroposterior condylar position at three time points: pre-surgery (T0), post-surgery (T1), and six months post-surgery (T2) using CBCT on 26 patients.

In the case of repeated measures ANOVA, if there is no existing data due to prior research, it is difficult to calculate the sample size or power. To calculate the power and sample size in Table 5 below, I generated data by setting hypothetical correlation coefficients for T0, T1, and T2 using the means and standard deviations from the table, and ran SPSS to obtain the results shown below "Fig. 8, Fig. 9". The Mauchly's test of sphericity satisfied the condition, yielding an epsilon of 1 for nonsphericity correlation. If sphericity is not satisfied, the Greenhouse-Geisser epsilon value is selected. Additionally, the variance explained by effect and error variance in G*power's From variances were obtained from the SStreat and SSE values in the SPSS results of the within-subject effects test table. The Number of Groups refers to the three time points T0, T1, and T2, so it is 3, and the Number of measurements is 26 patients at each time point, making it 26, while the Total sample size is 26*3=78. The variables Axial condylar angle and Anterior space (right) both had p-values less than 0.05 and a power greater than 0.8, indicating that a sufficient sample size was secured. However, to find significant differences in Superior space (right) and Posterior space (right), a larger sample size needs to be secured. If these variables are not the main outcome variables, there may be no need to secure additional sample sizes "Table 5".

## IV. Conclusion

Conducting clinical trials with a sample size of 30 without clinical or scientific evidence is not ethically or scientifically valid. Increasing the sample size always leads to statistically significant results[7]. This is because statistical significance is greatly influenced by sample size. For example, to statistically detect a very small effect size, 10,000 participants may be needed, while a sample size of just 30 may be sufficient to find a significant difference for a relatively large effect size. However, there are two scenarios where statistical significance may not be observed: first, when the sample size is insufficient, or second, when the actual effect size is smaller than expected. In such cases, the sample size calculated based on the expected effect size fails to reveal a significant difference. Even if the sample size is increased, a significant difference may be found, but if the actual effect size is too small, it may not have any clinical significance. Ultimately, the most important consideration is not to seek statistical significance, but to think about and study what effect size is clinically meaningful. Therefore, to conduct high-quality research, it is essential to determine the minimum clinically meaningful effect size during the planning stage of the study, decide on the sample size that can yield significant differences, and then begin the research.